\documentclass[groupedaddress,showpacs,showkeys,amssymb,eqsecnum,aps]{revtex4}
\usepackage[pdftex]{graphicx}
\usepackage{amsmath} 
\usepackage{epsf}                                                                                             
\usepackage{color}
\usepackage{verbatim}                                                                         
\newcommand{\be}{\begin{equation}}                                                                            
\newcommand{\ee}{\end{equation}}

\newcommand{\imineq}[2]{\vcenter{\hbox{\includegraphics[height=#2ex]{#1}}}}
\begin{document}                                                                                              
                                                                                                              
\title{Radiative Corrections and the Palatini Action}                                 
\author{F. T. Brandt}
\email{fbrandt@usp.br}
\affiliation{Instituto de F\'{\i}sica, Universidade de S\~ao Paulo, S\~ao Paulo, SP 05508-090, Brazil}
\author{D. G. C. McKeon}
\email{dgmckeo2@uwo.ca}
\affiliation{
Department of Applied Mathematics, The University of Western Ontario, London, ON N6A 5B7, Canada}
\affiliation{Department of Mathematics and Computer Science, Algoma University,
Sault St.Marie, ON P6A 2G4, Canada}

\date{\today}
                                                                                                              
\begin{abstract}                                                                                              
By using the Faddeev-Popov quantization procedure, we demonstrate that the radiative effects computed using the first-order and 
second-order Einstein-Hilbert action for General Relativity are the same, provided one can discard tadpoles. In addition, we show that 
the first order form of this action can be used to obtain a set of Feynman rules that involves just two propagating fields and three 
three-point vertices;  using these rules is considerably simpler than employing the infinite number of vertices that occur in the 
second-order form. We demonstrate this by computing the one-loop, two-point function. 
\end{abstract}                                                                                                
                                                                       
                                                               
\maketitle                     
\section{Introduction}
In the Einstein-Hilbert action
\begin{equation}\label{eq1}
S = \int d^d x \sqrt{-g} g^{\mu\nu} R_{\mu\nu}(\Gamma),
\end{equation}
where
\be \label{eq2}
\Gamma^\lambda_{\mu\nu} = \frac{1}{2} g^{\lambda\sigma} \left(
g_{\mu\sigma,\nu} + g_{\nu\sigma,\mu} - g_{\mu\nu,\sigma}
\right)
\ee
and
\be \label{eq3}
R_{\mu\nu}(\Gamma) =
\Gamma^\rho_{\mu\rho,\nu} - \Gamma^\rho_{\mu\nu,\rho} 
- \Gamma^\sigma_{\mu\nu}  \Gamma^\rho_{\sigma\rho} 
+\Gamma^\rho_{\mu\sigma}  \Gamma^\sigma_{\nu\rho} 
\ee
it is usual to take the metric $g_{\mu\nu}$ to be the independent variable and
the affine connection $\Gamma^\lambda_{\mu\nu} $ to be dependent; this is the second-order
Einstein-Hilbert action. Classically, it is possible to treat both $g_{\mu\nu}$ and $\Gamma^\lambda_{\mu\nu} $ 
as being independent;
the equation of motion for $\Gamma^\lambda_{\mu\nu} $  in this first order action yields Eq. \eqref{eq2}.
It was Einstein who first noted this, though the first-order Einstein-Hilbert (1EH) action is often attributed to Palatini \cite{Ferraris82}.

Although the 1EH and 2EH actions are equivalent at the classical level, it has as yet not been established that the two forms of the EH 
action result in the same quantum effects. We first show this quantum equivalence of the 1EH and 2EH actions when using the 
Faddeev-Popov procedure in conjunction with the quantum mechanical path integral, provided that tadpole integrals can 
be set equal to zero. This is of some consequence, as it has been noted \cite{McKeon:1994ds, Brandt:2015nxa} that the 
first order form of gauge theory actions is considerably simpler than the second order form. This is true both in Yang-Mills theory
(where two complicated vertices are replaced by one simple one that is independent of momentum) and in General Relativity (where a
single momentum independent vertex replaces an infinite series of momentum dependent vertices). The only disadvantage of using the first
order action is that there are now two propagating fields; in the 1EH case, these two fields have rather involved mixed propagator.

Our second result is that it is possible to shift variables of integration in the 1EH action within the path integral to eliminate this mixed
propagator. We are then left with a relatively simple set of Feynman rules; there are now just two propagating fields (that do not mix) and
three vertices. This is an improvement over the situation that occurs in the 2EH action where there is one propagating field and an infinite
number of vertices with an arbitrary number of external fields.

We then demonstrate the utility of our result by computing the two-point function to one-loop order using an arbitrary gauge fixing
parameter. In the limiting case in which this parameter equals one, we reproduce the result of Ref. \cite{Capper:1973pv}.

The first order formalism has also been used for doing loop calculations in gravity 
in Ref. \cite{Kalmykov:1994fm}, though in the models considered there it is not clear if the first and second order formalisms are equivalent.

%

We begin by considering the 1YM action.

\section{The first order Yang-Mills action}
It is evident that the 1YM Lagrangian
\be \label{eq4}
{\cal L}_{1YM} = - \frac 1 2 F^a_{\mu\nu}\left(\partial^\mu A^{a\, \nu} - \partial^\nu A^{a\, \mu} + g \epsilon^{abc} A^{b\, \mu} A^{c\, \nu}\right) 
+ \frac 1 4 F^a_{\mu\nu} F^{a\, \mu\nu}
\ee
is classically equivalent to the 2YM Lagrangian
\begin{equation}\label{eq5}
{\cal L}_{2YM} = - \frac 1 4 \left(\partial_\mu A^a_\nu - \partial_\nu A^a_\mu + 
g \epsilon^{abc} A^b_\mu A^c_\nu\right)^2  
\end{equation}
as upon substitution of the equation of motion for $F^a_{\mu\nu}$ that
follows from \eqref{eq4} back into ${\cal L}_{1YM}$, ${\cal L}_{2YM}$ follows.

The 1YM and 2YM Lagrangians have the gauge invariance
\begin{subequations}\label{eq6}
\begin{equation}\label{eq6a}
\delta F^a_{\mu\nu} =  g \epsilon^{abc} F_{\mu\nu}^b \theta^c 
\end{equation}
\begin{equation}\label{eq6b}
\delta A^a_\mu =\partial_\mu \theta^{a} + g \epsilon^{abc} A^b_\mu  \theta^c; 
\end{equation}
we are led to the path integral for ${\cal L}_{1YM}$
\be\label{eq7}
Z = \int {\cal D} A^a_\mu  {\cal D} F^a_{\mu\nu} 
\Delta_{FP}(A) 
\exp i\int d^d x\left( {\cal L}_{1YM}  + {\cal L}_{gf}\right) 
\ee
\end{subequations}
where $\Delta_{FP}(A)$ is the Faddeev-Popov determinant associated with 
the gauge fixing Lagrangian ${\cal L}_{gf}$. (More than one gauge fixing
may occur \cite{Brandt:2007td,Brandt:2009qi,McKeon:2014iea}.) The field
$A^a_\mu$  (but not $F^a_{\mu\nu}$) interacts with other ``matter'' fields.

If in Eq. \eqref{eq7} we perform the shift 
\begin{equation}\label{eq8}
 F^a_{\mu\nu} \rightarrow F^a_{\mu\nu} + \left(\partial_\mu A^a_\nu -\partial_\nu A^a_\mu 
+ g \epsilon^{abc} A^b_\mu A^c_\nu\right)  
\end{equation}
then we find that
\be\label{eq9}
Z = \int {\cal D} A^a_\mu  {\cal D} F^a_{\mu\nu} 
\Delta_{FP}(A) 
\exp i\int d^d x\left[
\frac 1 4 F^a_{\mu\nu} F^{a \mu\nu} + {\cal L}_{2YM}
+ {\cal L}_{gf}
\right] . 
\ee
the integral over $F^a_{\mu\nu}$ decouples and the usual generating functional
associated with ${\cal L}_{2YM}$ is recovered with its three-point and four-point
vertices. (In its unshifted form, Eq. \eqref{eq7} results in the three
propagators $\langle AA \rangle$, $\langle FF \rangle$ and $\langle AF \rangle$
and the vertex $\langle FAA \rangle$ \cite{McKeon:1994ds, Brandt:2015nxa}.)

We can also make the shift
\begin{equation}\label{eq10}
 F^a_{\mu\nu} \rightarrow F^a_{\mu\nu} 
+ \left(\partial_\mu A^a_\nu -\partial_\nu A^a_\mu \right)  
\end{equation}
leaving us with
\begin{eqnarray}\label{eq11}
Z &=& \int {\cal D} A^a_\mu  {\cal D} F^a_{\mu\nu} \Delta_{FP}(A) 
\exp i\int d^d x\left[ \frac 1 4 F^a_{\mu\nu} F^{a \mu\nu} 
- \frac 1 4 \left(\partial_\mu A^a_\nu -\partial_\nu A^a_\mu \right)^2  
\right. \nonumber \\ 
&&\left. \;\;\;\;\;\;\;\; - \frac 1 2 \left(F^a_{\mu\nu} + \partial_\mu A^a_\nu -\partial_\nu A^a_\mu \right)  
\left(g \epsilon^{abc} A^b_\mu A^c_\nu\right) 
+ {\cal L}_{gf}
\right] . 
\end{eqnarray}
When the generating functional $Z$ is written in this form we see that there are
now two propagators $\langle FF \rangle$ and $\langle AA \rangle$
as well as two three point functions $\langle FAA \rangle$ and $\langle AAA \rangle$
(but no mixed propagators $\langle AF \rangle$ or four point
vertex $\langle AAAA \rangle$.)

This possibility of altering the Feynman rules in YM theory will now be exploited
when examining the first order (Palatini) form of the Einstein-Hilbert action.

\section{The first order Einstein-Hilbert action}
Rather than using $g_{\mu\nu}$ and $\Gamma^\lambda_{\mu\nu}$ as independent fields
in the 1EH Lagrangian of Eq. \eqref{eq1}, it proves convenient to 
use \cite{McKeon:2010nf}
\begin{subequations}\label{eq12}
\be \label{eq12a}
h^{\mu\nu} = \sqrt{-g} g^{\mu\nu}
\ee
and
\be \label{eq12b}
G^\lambda_{\mu\nu} = \Gamma^\lambda_{\mu\nu} - \frac 1 2 \left(
\delta^\lambda_\mu \Gamma^\sigma_{\nu\sigma} + 
\delta^\lambda_\nu \Gamma^\sigma_{\mu\sigma} \right)
\ee
\end{subequations}
so that now we have 
\begin{equation}\label{eq13}
{\cal L}_{1EH} = h^{\mu\nu}\left(
G^\lambda_{\mu\nu\, ,\lambda}   +
\frac{1}{d-1} G^\lambda_{\mu\lambda}  G^\sigma_{\nu\sigma} -
G^\lambda_{\mu\sigma}  G^\sigma_{\nu\lambda} . 
\right) 
\end{equation}
The canonical structure of this action has been examined in refs.
\cite{McKeon:2010nf,Kiriushcheva:2009vh}
and the resulting path integral in ref. \cite{Chishtie:2012sq}.
Here, we will consider using the Faddeev-Popov path integral \cite{Faddeev:1967fc}
\be\label{eq14}
Z_{1EH} = \int {\cal D} h^{\mu\nu}  {\cal D} G^\lambda_{\mu\nu}
\Delta_{FP}(h) \exp i\int d^d x \left[{\cal L}_{1EH} + {\cal L}_{gf}\right]. 
\ee
Directly using the form of Eq. \eqref{eq13} makes it impossible
to define a propagator for $h^{\mu\nu}$ and $G^\lambda_{\mu\nu}$.
(This is easily seen if one were attempt to find a propagator for fields
$\phi$ and $V^\lambda$ with the Lagrangian ${\cal L} = \phi V^\lambda_{,\lambda}$.)
In ref. \cite{Capper:1973pv}, $h^{\mu\nu}$ is expanded about a flat metric
$\eta^{\mu\nu} = \mbox{diag}(+,+,+,\dots,-)$ so that
\be\label{eq15}
h^{\mu\nu}(x) = \eta^{\mu\nu} + \phi^{\mu\nu}(x);
\ee
the propagators 
$\langle \phi\phi \rangle$, $\langle GG \rangle$, $\langle \phi G \rangle$
and the vertex $\langle \phi G G \rangle$ are given in ref.  \cite{Brandt:2015nxa}.
However, it is not immediately evident how this form of $Z_{1EH}$ yields results
consistent with those that follow from the {2EH} Lagrangian
${\cal L}_{2EH}$.

To show this equivalence, we start by writing Eq. \eqref{eq13} as
\be\label {eq16}
{\cal L}_{1EH} =
G^\lambda_{\mu\nu}\left(- h^{\mu\nu}_{,\lambda}   \right)
+ \frac 1 2 M^{\mu\nu}_\lambda{}^{\pi\tau}_\sigma(h) 
G^\lambda_{\mu\nu} G^\sigma_{\pi\tau} ,
\ee
where 
\begin{eqnarray}\label{eq17}
M^{\mu\nu}_\lambda{}^{\pi\tau}_\sigma(h) & = &
\frac{1}{2}\left[\frac{1}{d-1}\left( \delta^\nu_\lambda\delta^\tau_\sigma h^{\mu\pi}+
                                                \delta^\mu_\lambda\delta^\tau_\sigma h^{\nu\pi}+
                                                \delta^\nu_\lambda\delta^\pi_\sigma h^{\mu\tau}+
                                                \delta^\mu_\lambda\delta^\pi_\sigma h^{\nu\tau}
\right) \right. 
\nonumber \\ 
&& - \left.\left( 
                                                \delta^\tau_\lambda\delta^\nu_\sigma h^{\mu\pi}+
                                                \delta^\tau_\lambda\delta^\mu_\sigma h^{\nu\pi}+
                                                \delta^\pi_\lambda\delta^\nu_\sigma h^{\mu\tau}+
                                                \delta^\pi_\lambda\delta^\mu_\sigma h^{\nu\tau}
\right) \frac{}{} \!\!\right]
\end{eqnarray}
From Eq. \eqref{eq16} we obtain the equation of motion
\be\label{eq18}
h^{\mu\nu}_{,\lambda} = 
M^{\mu\nu}_\lambda{}^{\pi\tau}_\sigma(h) G^\sigma_{\pi\tau}
\ee
from which we see that (upon using Eq. \eqref{eq17} and with 
$h_{\mu\lambda} h^{\lambda\nu} = \delta_\mu^\nu$)
\begin{eqnarray}\label{eq19}
H_{\pi\tau,\lambda} & \equiv &
-h_{\pi\mu} h_{\tau\nu} h^{\mu\nu}_{,\lambda} 
+h_{\tau\mu} h_{\lambda\nu} h^{\mu\nu}_{,\pi}  
+h_{\lambda\mu} h_{\pi\nu} h^{\mu\nu}_{,\tau} 
\nonumber \\ & = &
2\left(
\frac{1}{d-1} h_{\pi\tau} G^\sigma_{\lambda\sigma}
-h_{\lambda\sigma} G^\sigma_{\pi\tau}
\right) .
\end{eqnarray}
Upon contracting Eq. \eqref{eq19} with $h^{\tau\lambda}$ we see that
\be\label{eq20}
G^\sigma_{\pi\sigma} = - \frac{d-1}{2(d-2)} h_{\mu\nu} h^{\mu\nu}_{,\pi}
\ee
and so by Eq. \eqref{eq19}
\be\label{eq21}
G^\rho_{\pi\tau} = \frac 1 2 h^{\rho\lambda}\left(
-\frac{1}{d-2} h_{\pi\tau} h_{\mu\nu} h^{\mu\nu}_{,\lambda}
-H_{\pi\tau,\lambda}
\right).
\ee
From Eq. \eqref{eq19} it is apparent that
\begin{eqnarray}\label{eq22}
\left(M^{-1}\right){}^\rho_{\pi\tau}{}^\lambda_{\mu\nu}(h) &=& 
\frac{-1}{2(d-2)}h^{\rho\lambda} h_{\pi\tau} h_{\mu\nu} +
\frac 1 4 h^{\rho\lambda}  \left(h_{\pi\mu} h_{\tau\nu} + h_{\pi\nu} h_{\tau\mu} \right) 
\nonumber\\ &-&
\frac 1 4 \left(h_{\tau\mu} \delta^\rho_\nu  \delta^\lambda_\pi +
h_{\pi\mu} \delta^\rho_\nu  \delta^\lambda_\tau +
h_{\tau\nu} \delta^\rho_\mu  \delta^\lambda_\pi  +
h_{\pi\nu} \delta^\rho_\mu  \delta^\lambda_\tau\right)  
\end{eqnarray}
(We have
\be\label{eq23}
\left. 
\left(M^{-1}\right){}^\rho_{\alpha\beta}{}^\lambda_{\mu\nu} \,
M^{\mu\nu}_\lambda {}^{\gamma\delta}_\sigma = 
\Delta^{\gamma\delta}_{\alpha\beta} \delta^\rho_\sigma \equiv
\frac{1}{2}\left(
\delta^\gamma_\alpha\delta^\delta_\beta  +
\delta^\delta_\alpha\delta^\gamma_\beta 
\right)\delta^\rho_\sigma 
\right)
\ee

In the Lagrangian of Eq. \eqref{eq16} we insert Eq. \eqref{eq21} and obtain
\be\label{eq24}
{\cal L}_{1EH} = -\frac 1 2 h^{\mu\nu}_{,\lambda}
\left(M^{-1}\right){}^\lambda_{\mu\nu}{}^\sigma_{\pi\tau}(h) h^{\pi\tau}_{,\sigma}
\ee
which is just the second-order EH Lagrangian ${\cal L}_{2EH}$. This
demonstrates that classically, ${\cal L}_{1EH}$ and ${\cal L}_{2EH}$ are equivalent.

We now make the shift
\be\label{eq25}
G^\lambda_{\mu\nu} \rightarrow G^\lambda_{\mu\nu} +
\left(M^{-1}\right){}^\lambda_{\mu\nu}{}^\sigma_{\pi\tau}(h) h^{\pi\tau}_{,\sigma}
\ee
in the path integral of Eq. \eqref{eq14}. We then find that
\be\label{eq26}
Z_{1EH} = \int {\cal D} h^{\mu\nu} {\cal D} G^\lambda_{\mu\nu}
\Delta_{FP}(h) \exp i\int d^d x \left[\frac 1 2 
G^\lambda_{\mu\nu} M^{\mu\nu}_\lambda{}^{\pi\tau}_\sigma(h) G_{\pi\tau}^\sigma
+\frac 1 2 h^{\mu\nu}_{,\lambda} \left(M^{-1}\right){}^\lambda_{\mu\nu}
{}^\sigma_{\pi\tau}(h) h^{\pi\tau}_{,\sigma} + {\cal L}_{gf}\right].
\ee
The expansion of Eq. \eqref{eq15} can now be made in Eq. \eqref{eq26}. Since
$M$ is linear in $h^{\mu\nu}$, it follows that
\be\label{eq27}
M^{\mu\nu}_\lambda{}^{\pi\tau}_\sigma(\eta+\phi) = 
M^{\mu\nu}_\lambda{}^{\pi\tau}_\sigma(\eta) +
M^{\mu\nu}_\lambda{}^{\pi\tau}_\sigma(\phi).  
\ee
Consequently, any Feynman diagrams contributing to Green's functions with only
the field $\phi^{\mu\nu}$ on external legs and which involve the field
$G^\lambda_{\mu\nu}$ on internal lines, necessarily will have the field 
$G^\lambda_{\mu\nu}$  appearing in a closed loop. But the propagator for
the field $G^\lambda_{\mu\nu}$  is independent of momentum (see Eq. \eqref{eq22})
and hence the loop momentum integral associated with any loop
coming from the field $G^\lambda_{\mu\nu}$  is of the form
\be\label{eq28}
\int d^d k P(k^\mu),
\ee
where $P(k^\mu)$ is a polynomial in the loop momentum $k^\mu$. If we use
dimensional regularization  \cite{tHooft:1972fi,Leibbrandt:1975dj} then such
loop momentum integrals vanish.

Consequently, for Green's functions involving only the field $\phi^{\mu\nu}$ on external
legs, the only contribution to Feynman diagrams come from the last two terms in the
argument of the exponential in Eq. \eqref{eq26}; from Eq. \eqref{eq24} we see that
this is just the generating functional associated with $-{\cal L}_{2EH}$ and so these
Green's functions can be derived by using either the first order or the second order
form of the EH action.

Using the second order form with the Lagrangian of Eq. \eqref{eq24} results in an
infinite series of vertices involving the field $h^{\mu\nu}$ 
(see ref. \cite{Capper:1973pv}). To obtain them, we note that when Eq. \eqref{eq27}
is substituted into Eq. \eqref{eq23}, we schematically obtain
\be\label{eq29}
\left(M^{-1}\right)(\eta+\phi) =
{M}^{-1}(\eta) - M^{-1}(\eta) M(\phi) M^{-1}(\eta) +
M^{-1}(\eta) M(\phi) M^{-1}(\eta) M(\phi) M^{-1}(\eta) - \dots .
\ee
The first term in Eq. \eqref{eq29} is associated with the propagator for
the $\phi^{\mu\nu}$ field  in the second order formalism while each subsequent term
is associated with a vertex. This means that direct use of the 2EH Lagrangian
becomes exceedingly complicated if more than the one-loop two-point
Green's function is to be computed \cite{tHooft:1974bx,Goroff:1985sz}.

We now will show that the 1EH generating functional can be used to compute
Green's functions with only the two propagators 
$\langle \phi\phi \rangle$, $\langle GG \rangle$ and the three point functions
$\langle GG\phi \rangle$, $\langle G\phi\phi \rangle$ and 
$\langle \phi\phi\phi \rangle$. First the expansion of Eq. \eqref{eq15} is
made and then the shift occurs
\be\label{eq30}
G^\lambda_{\mu\nu} \rightarrow G^\lambda_{\mu\nu} +
\left(M^{-1}\right){}^\lambda_{\mu\nu}{}^\sigma_{\pi\tau}(\eta) h^{\pi\tau}_{,\sigma}
\ee
(This is the shift of Eq. \eqref{eq25} with $h$ being replaced by $\eta$.) This leads
to Eq. \eqref{eq14} becoming
\begin{eqnarray}\label{eq31}
Z_{1EH} &=& \int {\cal D} h^{\mu\nu}  {\cal D} G^\lambda_{\mu\nu}
\Delta_{FP}(h) \exp i\int d^d x\left[
\frac 1 2 G^\lambda_{\mu\nu} M^{\mu\nu}_\lambda{}^{\pi\tau}_\sigma(\eta) G^\sigma_{\pi\tau}
-\frac 1 2 \phi^{\mu\nu}_{,\lambda} {M^{-1}}{}_{\mu\nu}^\lambda{}^\sigma_{\pi\tau}(\eta) 
\phi^{\pi\tau}_{,\sigma} 
\right. \nonumber \\ && \left.
+ \frac 1 2\left(
G^\lambda_{\mu\nu}+\phi^{\alpha\beta}_{,\rho}
\left(M^{-1}\right){}^\rho_{\alpha\beta}{}^\lambda_{\mu\nu}(\eta)
\right)
\left(M^{\mu\nu}_\lambda{}^{\pi\tau}_\sigma(\phi)\right)
\left(
G^\sigma_{\pi\tau}+\left(M^{-1}\right){}^\sigma_{\pi\tau}{}^\xi_{\gamma\delta}(\eta)
\phi^{\gamma\delta}_{,\xi}
\right) + {\cal L}_{gf}
\right] . 
\end{eqnarray}
The contributions coming from the various terms in the argument of the exponential
appearing in Eq. \eqref{eq31} that lead to the Feynman rules can be immediately 
seen to be:
\begin{subequations}\label{eq32}
\be\label{eq32a}
G\mbox{-}G : \frac 1 2 G^\lambda_{\mu\nu} 
M^{\mu\nu}_\lambda{}^{\pi\tau}_\sigma(\eta)
                      G^\sigma_{\pi\tau} :
\ee
\be\label{eq32b}
\phi\mbox{-}\phi : -\frac 1 2 \phi^{\mu\nu}_{,\lambda} 
{M^{-1}}{}_{\mu\nu}^\lambda{}^\sigma_{\pi\tau}(\eta) 
\phi^{\pi\tau}_{,\sigma}  - \frac{1}{2\alpha}(\phi^{\mu\nu}_{,\nu})^2:
\ee
\be\label{eq32c}
G\mbox{-}G\mbox{-}\phi:\frac 1 2 M^{\mu\nu}_\lambda{}^{\pi\tau}_\sigma(\phi) 
G^\lambda_{\mu\nu} G^\sigma_{\pi\tau} :
\ee
\be\label{eq32d}
G\mbox{-}\phi\mbox{-}\phi : G^\lambda_{\mu\nu} 
M^{\mu\nu}_\lambda{}^{\pi\tau}_\sigma(\phi) 
{M^{-1}}^\sigma_{\pi\tau}{}_{\gamma\delta}^\xi(\eta) \phi^{\gamma\delta}_{,\xi} :
\ee
\be\label{eq32e}
\phi\mbox{-}\phi\mbox{-}\phi : \frac 1 2 \phi^{\alpha\beta}_{,\rho}
{M^{-1}}^\rho_{\alpha\beta}{}^\lambda_{\mu\nu}(\eta) 
M^{\mu\nu}_\lambda{}^{\pi\tau}_\sigma(\phi) 
{M^{-1}}^\sigma_{\pi\tau}{}^\xi_{\gamma\delta}(\eta) 
\phi^{\gamma\delta}_{,\xi} : .
\ee
\end{subequations}
In Eq. \eqref{eq32b} we have used the gauge fixing Lagrangian
\be\label{eq33}
{\cal L} = -\frac{1}{2\alpha}(\phi^{\mu\nu}_{,\nu})^2.
\ee
With this gauge fixing, the contribution coming from the Faddeev-Popov
determinant $\Delta_{FP}$ in Eq. \eqref{eq31} involves the Feynman rules
that follow from \cite{Capper:1973pv,Brandt:2015nxa}
\begin{eqnarray}\label{eq34}
{\cal L}_{ghost} &=& \bar d_\mu\left[ \partial^2 \eta^{\mu\nu} +
(\phi^{\rho\sigma}_{,\rho})\partial_\sigma \eta^{\mu\nu}-
(\phi^{\rho\mu}_{,\rho})\partial^\nu \right. \nonumber \\
& &  \;\;\;\;\;\;\;\;\;\;\;  \left.  + \phi^{\rho\sigma} \partial_\rho\partial_\sigma \eta^{\mu\nu} -(\partial_\rho\partial^\nu\phi^{\rho\mu}) \right]d_\nu ,
\end{eqnarray}
where $d^\mu$ and $\bar d^\mu$ are Fermionic vector ghost fields.
These are found to be
\begin{subequations}\label{eq35}
\be\label{eq35a}
\bar d\mbox{-}d :\bar d_\mu\partial^2 d_\nu :
\ee
\be\label{eq35b}
\bar d\mbox{-}d\mbox{-}\phi 
:  \bar d_\mu\left[
(\phi^{\rho\sigma}_{,\rho})\partial_\sigma \eta^{\mu\nu}-
(\phi^{\rho\mu}_{,\rho})\partial^\nu 
+ \phi^{\rho\sigma} \partial_\rho\partial_\sigma \eta^{\mu\nu} 
-(\partial_\rho\partial^\nu\phi^{\rho\mu}) \right]d_\nu :
\ee
\end{subequations}

Let us now consider an explicit calculation of an one-loop radiative correction.
From Eqs. \eqref{eq32} and \eqref{eq35} we readily find the 
following momentum space Feynman rules (all vertex momenta are inwards and $p+q+r=0$)
\begin{subequations}\label{feynrules}
\begin{eqnarray}
\displaystyle{\substack{{}^{\displaystyle{ {}_{\mu\, \nu}}} 
{\; \includegraphics[scale=0.7]{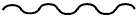} \;} {}^{\displaystyle{{}_{\rho\sigma}}} 
\\  {\displaystyle{p}}}}
&:& \;\;\;\;
\frac{(1- \alpha) \left(p^{\nu } p^{\sigma } \eta^{\mu \rho }+p^{\nu 
   } p^{\rho } \eta^{\mu \sigma }+p^{\mu } p^{\sigma } \eta^{\nu \rho 
   }+p^{\mu } p^{\rho } \eta^{\nu \sigma } -2 p^{\rho } p^{\sigma } \eta^{\mu \nu } -2 p^{\mu } p^{\nu } \eta^{\rho \sigma }  \right)}{p^4} 
\nonumber 
\\
& & 
-\frac{\eta^{\mu \sigma } \eta^{\nu \rho }+\eta^{\mu \rho } \eta^{\nu \sigma }
-(2-\alpha)  \eta^{\mu \nu } \eta^{\rho \sigma }}{p^2}
%
\end{eqnarray}
\begin{eqnarray}
\displaystyle{\substack{{}^{\displaystyle{ {}_{\mu\nu}^\lambda}} {\; \includegraphics[scale=0.7]{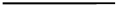} \;} {}^{\displaystyle{ {}_{\pi\tau}^\rho}} 
\\  {\displaystyle{ }}}} 
&:& \;\;\;\;
%
%
\frac{1}{4} \eta^{\lambda \rho } \left(\eta_{\mu\tau } \eta_{\nu \pi} + \eta_{\mu \pi } \eta_{\nu \tau}
-\frac{2}{d-2} \eta_{\mu \nu } \eta_{\pi \tau }\right) 
\nonumber \\
&&
-\frac{1}{4} \left( \delta^{\lambda}_{ \tau } \delta_{\mu}^{ \rho } \eta_{\nu \pi}
+\delta^{\lambda}_{\tau } \delta_{\nu}^{ \rho}\eta_{\mu \pi } 
+\delta^{\lambda}_{  \pi }  \delta_{\nu}^{ \rho} \eta_{\mu \tau } 
+\delta^{\lambda}_{ \pi } \delta_{\mu}^{ \rho } \eta_{\nu \tau} \right) 
\equiv {\cal D}{}_{\mu\nu}^\lambda{}_{\pi\tau}^\rho  
%
\end{eqnarray}
\begin{eqnarray}
{}_{\mu\nu} { \imineq{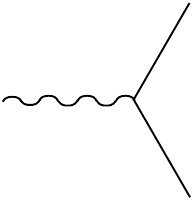}{14} }^{\displaystyle{\;\;{}^{\alpha\beta}_\lambda}}_{\displaystyle{\;\;{}^{\gamma\delta}_\sigma}}&:& \;\;\;\;
\frac{1}{8} \left\{\left[\left(
\frac{\delta_{\mu}^{\beta} \delta_{\nu}^{\delta} \delta^\alpha_\lambda \delta^\gamma_\sigma}{d-1} - 
\delta_{\mu}^{\beta} \delta_{\nu}^\delta \delta^\alpha_\sigma \delta^\gamma_\lambda + \mu \leftrightarrow \nu\right) + \alpha\leftrightarrow\beta 
\right]+\gamma\leftrightarrow\delta\right\}
\nonumber \\
&& + \;\; (\lambda, \alpha,\beta) \longleftrightarrow (\sigma,\gamma,\delta)
\end{eqnarray}
\begin{eqnarray}
{}_\sigma^{\gamma\delta}\, p\, { \imineq{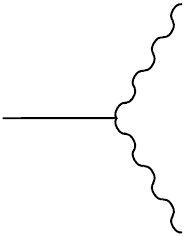}{14} }^{\displaystyle{q\;{}_{\mu\nu}}}_{\displaystyle{r\;{}_{\alpha\beta}}}
&:& \;\;\;\;
\frac{i r_\theta}{4}\left\{\left[\left(
\frac{1}{d-1} \delta^\gamma_\mu \delta^\delta_\sigma {\cal D}_{ \alpha\beta}^\theta {}^\rho_{\nu\rho}
-\delta^\gamma_\mu {\cal D}_{\alpha\beta}^\theta{}_{\nu\sigma}^\delta  
+ \mu \leftrightarrow \nu\right) + \alpha\leftrightarrow \beta\right] + \gamma\leftrightarrow \delta\right\}
\nonumber \\
&& +  \;\; (q, \alpha,\beta) \longleftrightarrow (r,\mu,\nu)
\end{eqnarray}
\begin{eqnarray}
{}_{\mu\nu}\, p\, { \imineq{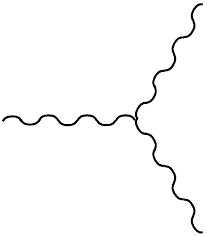}{14} }^{\displaystyle{q\;{}_{\alpha\beta}}}_{\displaystyle{r\;{}_{\gamma\delta}}}
&:& \;\;\;\;
\frac{q_\kappa r_\theta}{8}\left\{\left[\left(
{\cal D}{}^\kappa_{\alpha\beta}{}^\pi_{\mu\sigma} 
                                                                {\cal D}{}^\theta_{\gamma\delta}{}^\sigma_{\nu\pi} 
 -\frac{1}{d-1} {\cal D}{}^\kappa_{\alpha\beta}{}^\sigma_{\mu\sigma} 
                                                                {\cal D}{}^\theta_{\gamma\delta}{}^\pi_{\nu\pi} 
 + \mu \leftrightarrow \nu\right) + \alpha\leftrightarrow \beta\right] + \gamma\leftrightarrow \delta\right\}
\nonumber \\
&& + \;\; \mbox{six permutations of}\;\; (p, \mu,\nu) \;\; (q, \alpha,\beta) \;\; (r, \gamma,\delta) 
\end{eqnarray}
\begin{eqnarray}
\substack{ {}^{\displaystyle{\bar d_\mu }}\; \includegraphics[scale=0.7]{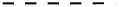} \; {}^{\displaystyle{ d_\nu }}
\\  {\displaystyle{ p}}}
&:& \;\;\;\;
-\frac{\eta^{\mu\nu}}{p^2}
\end{eqnarray}
\begin{eqnarray}
{}_{\mu\nu}\, p\, { \imineq{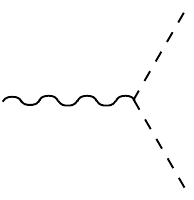}{14} }^{{\displaystyle{q\;{}_\alpha}}}_{{{\displaystyle{r {}_\beta\;\; }}}}
&:& \;\;\;\;
\frac{1}{2}\left[
\eta_{\alpha\beta}\left(q_\mu r_\nu + q_\nu r_\mu\right) 
-q_\beta\left(p_\mu \eta_{\alpha\nu} + p_\nu \eta_{\alpha\mu}\right) 
\right]
\end{eqnarray}
\end{subequations}

The one-loop contributions to the two-point function $\langle \phi\phi \rangle$ are given
by the Feynman diagrams of fig. 1. 
After loop integration,  the result can only depend (by covariance) on the five tensors shown 
in table I, so that each diagram in figure 1 can be written as 
\begin{equation}\label{eq2a}
\Pi^I_{\mu\nu\,\alpha\beta}(k) =  \sum_{i=1}^{5}  {\cal T} ^i_{\mu\nu\, \alpha\beta}(k) 
C^I_i(k) ; \;\;\; I=\mbox{a, }\mbox{b, }\mbox{c and }\mbox{d} .
\end{equation}
The coefficients $C^I_i$ can be obtained solving the following system of five algebraic equations
\be
\sum_{i=1}^5 {\cal T}^i_{\mu\nu\,\alpha\beta}(k) {\cal T}^j{}^{\mu\nu\,\alpha\beta}(k) C^I_i(k) =
\Pi^I_{\mu\nu\alpha\beta}(k) {\cal T}^j{}^{\mu\nu\,\alpha\beta}(k) \equiv J^I{}^j(k); \;\;
j=1,\dots,5 .
\ee
Using the Feynman rules for $\Pi^I_{\mu\nu\,\alpha\beta}(k)$ 
the integrals on the right hand side have the following form
\be
J^I{}^j(k) = \int \frac{d^d p}{(2 \pi)^d}  s^I{}^j(p,q,k).
\ee
where $q=p+k$; $p$ is the loop momentum, $k$ is the external momentum and $s^I{}^j(p,q,k)$ are
scalar functions. Using the relations 
\begin{subequations}
\begin{eqnarray}
p\cdot k = (q^2 - p^2 - k^2)/2, \\
q\cdot k = (q^2 + k^2 - p^2)/2, \\
p\cdot q = (p^2 + q^2 - k^2)/2, 
\end{eqnarray}
\end{subequations}
the scalars  $s^I{}^j(p,q,k)$ can be reduced to combinations of powers of $p^2$ and $q^2$. As a result, the
integrals $J^I{}^j(k)$  can be expressed in terms of combinations of the following well known integrals 
\begin{equation}
I^{ab} \equiv 
\int \frac{d^d p}{(2 \pi)^d} \frac{1}{(p^2)^a (q^2)^b} = \frac{(k^2)^{d/2-a-b}}{(4\pi)^{d/2}}
\frac{\Gamma(a+b-d/2)}{\Gamma(a) \Gamma(b)} \frac{\Gamma(d/2-a) \Gamma(d/2-b)}{\Gamma(d-a-b)} 
\end{equation}
(this has also been considered in \cite{Chetyrkin:1980pr}).
The only non-vanishing (ie non tadpole) integrals are the ones with both $a> 0$ and $b> 0$. 
As we have pointed out earlier the integrals $J^{\mbox{a}}_i(k)$ and $J^{\mbox{b}}_i(k)$, 
associated respectively with the diagrams (a) and (b) of figure 1, are tadpole like and will not contribute
%
(either $a$ or $b$ is not positive). For a general gauge parameter, $\alpha \neq 1$ 
the diagram (c) in figure 1 involves the following three kinds of integrals
\begin{subequations}\label{intregd}
\begin{eqnarray}
I^{11} & = & \frac{(k^2)^{d/2-2}}{2^d\pi^{d/2}}
\frac{\Gamma \left(2-\frac{d}{2}\right) \Gamma \left(\frac{d}{2}-1\right)^2}{\Gamma (d-2)} \\
I^{12} & = & I^{21} = \frac{(3-d) I^{11}} {k^2} \\ 
I^{22} & = & \frac{(3-d) (6-d) I^{11} } {k^4} .
\end{eqnarray}
\end{subequations}
The ghost loop diagram only involves $I^{11}$.

A straightforward computer algebra code can now be setup in order implement the steps above described
and to obtain the structures
$C^{\mbox{c}}_i$ and $C^{\mbox{d}}_i$. The results are the following
\begin{subequations}
\begin{eqnarray}
C^{\mbox{c}}_1 &=& \frac{1}{8(d-1)}\left[
\frac{1}{8} \left(d^3-2 d^2+96 d-64\right)
-4 (\alpha-1)    (2 d^2-11 d+8) 
\right.\nonumber \\ &+& \left.
2(\alpha-1)^2   (d-1) (d^2-6 d+12)   
\right] I^{11}
\end{eqnarray}
\begin{eqnarray}
C^{\mbox{c}}_2 &=& \frac{1}{8(d-1)(d-2)^2}\left[
\frac{1}{8} d \left(-7 d^2+4 d+52\right) 
+4(\alpha-1)    (d^3 - 9 d^2 + 23 d - 14)   
\right.\nonumber \\ &-& \left.
(\alpha-1)^2    (d - 4) (d - 1) (d^2 - 7 d + 14) 
\right] k^4 I^{11}
\end{eqnarray}
\begin{eqnarray} 
C^{\mbox{c}}_3 &=& \frac{1}{32(d-1)}\left[
\frac{1}{2} \left(4 d^2+5 d-16\right)
-8(\alpha-1)    (d - 5) (d - 1)   
\right.\nonumber \\ &+& \left.
2 (\alpha-1)^2   (d - 1) (d^2 - 7 d + 16)   
\right]k^4 I^{11}
\end{eqnarray}
\begin{eqnarray} 
C^{\mbox{c}}_4 &=& \frac{1}{16(d-1)(d-2)}\Bigl[
\frac{1}{4} \left(d^3-2 d^2+40 d+16\right)
-4 (\alpha-1) (3 d^2-17 d+12)  
\nonumber \\ &+& 
4 (\alpha-1)^2  (d - 1) (d^2 - 6 d + 12) 
\Bigr] k^2 I^{11}
\end{eqnarray}
\begin{eqnarray} 
C^{\mbox{c}}_5 &=& \frac{1}{32(d-1)}\left[
\frac{1}{2} \left(-4 d^2-5 d+20\right)
+8 (\alpha-1)   (d - 5) (d - 1) 
\right.\nonumber \\ &-& \left.
2(\alpha-1)^2   (d - 1) (d^2 - 7 d + 16)   
\right] k^2 I^{11}
\end{eqnarray}
\end{subequations}

\begin{subequations}
\be 
C^{\mbox{d}}_1 = -\frac{(d-2) \left(d^2+8 d+8\right)}{16 (d^2-1) } I^{11}
\ee 
\be 
C^{\mbox{d}}_2 = C^{\mbox{d}}_3 = -\frac{d }{16 (d^2-1) } k^4 I^{11}
\ee 
\be 
C^{\mbox{d}}_4 = -\frac{\left(d^2+2 d+2\right) }{16 (d^2-1) } k^2 I^{11}
\ee 
\be 
C^{\mbox{d}}_5 = -\frac{1}{16 (d^2-1) } k^2 I^{11}
\ee 
\end{subequations}
The final expression for the one-loop contribution to $\langle \phi\phi \rangle$ can now be expressed as
\be
\Pi_{\mu\nu\,\alpha\beta} = \sum_{i=1}^5 \left(C^{\mbox{c}}_i + C^{\mbox{d}}_i\right) {\cal T}^i_{\mu\nu\,\alpha\beta}
= I^{11} \sum_{i=1}^5 C_i  {\cal T}^i_{\mu\nu\,\alpha\beta} ,
\ee
where
\begin{subequations}\label{eq35n}
  \be
   C_1 =  \frac{1}{4} (\alpha -1)^2 \left(d^2-6 d+12\right)-\frac{(\alpha -1)
   \left(2 d^2-11 d+8\right)}{2 (d-1)}+\frac{d \left(d^3-5 d^2+70
   d+64\right)}{64 (d-1) (d+1)},
   \ee
     \be
   C_2 = \left[-\frac{(\alpha -1)^2 (d-4) \left(d^2-7 d+14\right)}{8
       (d-2)^2}
+\frac{(\alpha -1) \left(d^3-9 d^2+23 d-14\right)}{2 (d-2)^2 (d-1)}
     -\frac{(d-3) d \left(7 d^2+28 d+12\right)}{64 (d-2)^2 (d-1)
       (d+1)}
          \right] k^4,
   \ee
     \be
     C_3 =  \left[\frac{1}{16} (\alpha -1)^2 \left(d^2-7 d+16\right)
-\frac{1}{4} (\alpha -1) (d-5)
     +\frac{4 d^3+9 d^2-15 d-16}{64 (d-1) (d+1)}\right] k^4 ,
   \ee
     \be\label{35d}
   C_4 = \frac{k^{2} {C_1}}{2 (d-2)}+\frac{(d-2) {C_2}}{2 k^{2}}+\frac{{C_3}}{k^{2}} 
   \ee
    \mbox{and} 
     \be\label{35e}
   C_5 = -\frac{C_3}{k^2}.
  \ee
\end{subequations}
Eqs. \eqref{35d} and \eqref{35e} are a consequence of the Ward identity
\be
(\eta^{\mu\rho} k^\nu + \eta^{\nu\rho} k^\mu - \eta^{\mu\nu} k^{\rho})
  \Pi_{\mu\nu\,\alpha\beta} 
(\eta^{\alpha\sigma} k^\beta + \eta^{\beta\sigma} k^\alpha - \eta^{\alpha\beta} k^{\sigma}) = 0
  \ee
which follows from the diffeomorphism invariance of the Einstein-Hilbert action \cite{Brandt:2015nxa}.
In the special case when $\alpha=1$ Eqs. \eqref{eq35n} are in agreement with ref. \cite{Capper:1973pv}.

\begin{table}[htbp]
\begin{center}
\begin{tabular}{c l}\hline \hline \\
${\cal T}^{1} _{\mu \nu \,\alpha \beta} (u,k)=$&$  k_\mu k_\nu k_\alpha  k_\beta$\\ & \\
${\cal T}^2 _{\mu \nu \,\alpha \beta} (u,k)=$&$ \eta_{\mu \nu} \eta_{\alpha\beta}$ \\ & \\
${\cal T}^3 _{\mu \nu \, \alpha \beta}(u,k)=$&$  \eta _{\mu \alpha} \eta_{\nu \beta} + \eta _{\mu \beta} \eta_{\nu \alpha}$  \\ & \\
${\cal T}^4 _{\mu \nu \,\alpha \beta} (u,k)=$&$ \eta_{\mu \nu} k_\alpha k_\beta + \eta_{\alpha \beta} k_\mu k_\nu$ \\ & \\
${\cal T}^5 _{\mu \nu \, \alpha \beta}(u,k)=$&$ \eta_{\mu \alpha} k_\nu k_\beta +
\eta_{\mu \beta} k_\nu k_\alpha + \eta_{\nu \alpha} k_\mu k_\beta +
\eta_{\nu \beta} k_\mu k_\alpha$ 
\\  \\ \hline \hline 
\end{tabular}\caption{The five independent tensors built from 
$\eta_{\mu\nu}$ and $k_\mu$, satisfying the symmetry conditions 
${\cal T}^{i}_{\mu\nu\,\alpha\beta} (k)= {\cal T}^{i}_{\nu\mu\,\alpha\beta}(k)= 
{\cal T}^{i}_{\mu\nu\,\beta\alpha} (k)= {\cal T}^{i}_{\alpha\beta\,\mu\nu}(k)$.}
\end{center}
\end{table}\label{tab1}

%

\begin{figure}[ht]\label{fig1}
\begin{eqnarray}
\includegraphics[scale=0.6]{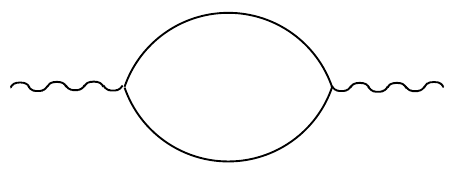} \atop \mbox{(a)} & 
\includegraphics[scale=0.6]{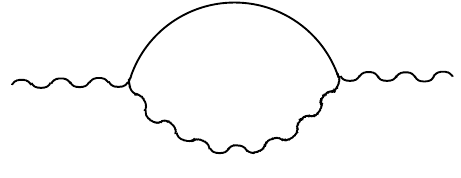} \atop \mbox{(b)}  
\nonumber 
\\
\nonumber 
\\
\includegraphics[scale=0.6]{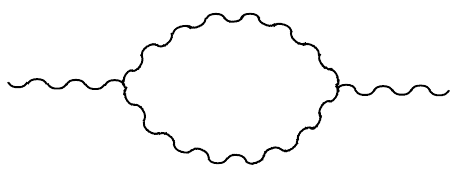} \atop \mbox{(c)} & 
\includegraphics[scale=0.6]{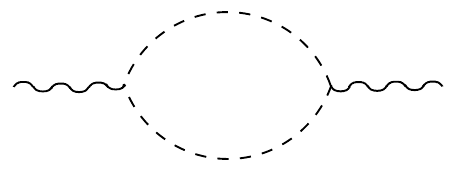} \atop \mbox{(d)}  
\nonumber 
\end{eqnarray}
\caption{One-loop contributions to $\langle \phi\phi \rangle$.}
\label{fig3}
\end{figure}

\newpage

\section{Discussion}
Establishing the equivalence between the first and second order forms of the
Yang-Mills Lagrangians at both the classical and quantum levels is straightforward;
this was demonstrated in section two above. It is not so easy to show at both the
classical and quantum levels that the first- and second-order forms of the 
Einstein-Hilbert action are equivalent. In section three above we have shown that
this equivalence holds provided it is possible to discard tadpole diagrams (which
are regulated to zero when using dimensional regularization.) (One feature
of this demonstration whose significance is not immediately apparent is the
difference in sign between ${\cal L}_{1EH}$ in \eqref{eq24} and the
$h M^{-1}(h) h$ term in Eq. \eqref{eq26}.)

We have also shown that by rewriting the 1EH action judiciously, it is possible to
have just two propagating fields and three three-point functions. This may prove
to be an advantage when considering higher order diagrams in the loop expansion
in (super-)gravity.

It is quite straightforward to adopt the methods of refs. 
\cite{tHooft:1974bx,Goroff:1985sz,vandeVen:1992gw,Mann:1988xy},
involving the use of geodesic coordinates in conjunction with a background field
for $\phi^{\mu\nu}$, to determine counter terms while working with
the 1EH Lagrangian.

It would also be interesting to compute the one loop correction to the two-point function
$\langle \phi\phi \rangle$ using the transverse-traceless (TT) gauge of ref. \cite{Brandt:2007td}.

\begin{acknowledgments}
We would like to thank CNPq and Fapesp (Brazil) for financial support.
Profs. J. Frenkel and J. C. Taylor have made helpful comments on this manuscript. 
An enlightening conversation with Roger Macleod is acknowledged.
\end{acknowledgments}

\newpage


\end{document}